\documentstyle[aps,twocolumn]{revtex}
\def\d{d}

\def\o{\otimes}
\def\half{{\textstyle \frac{1}{2}}}
\def\g4{\hat g}
\def\r2{2^{\half}}

\begin{document}
\twocolumn[\hsize\textwidth\columnwidth\hsize\csname @twocolumnfalse\endcsname

\title{Hidden symmetry of the three-dimensional Einstein--Maxwell equations}
\author{Daisuke Ida $^{\dag}$ and Yoshiyuki Morisawa $^{\ddag}$}
\address{$^{\dag}$ Department of Physics, Kyoto University,
Kyoto 606-8502, Japan\\ 
{\rm E-mail: ida@tap.scphys.kyoto-u.ac.jp}\\
$^{\ddag}$ Yukawa Institute for Theoretical Physics,
Kyoto University,
Kyoto 606-8502, Japan\\ 
{\rm E-mail: morisawa@yukawa.kyoto-u.ac.jp}\\
}
\date{January 29, 2001}
\maketitle
\begin{abstract}
It is shown how to generate three-dimensional Einstein--Maxwell fields from known ones
in the presence of a hypersurface-orthogonal non-null Killing vector field.
The continuous symmetry group is isomorphic to the Heisenberg group including the
Harrison-type transformation.
The symmetry of the Einstein--Maxwell--dilaton system is also studied and
it is shown that there is the $SL(2,{\bf R})$ transformation between the Maxwell and
the dilaton fields. This $SL(2,{\bf R})$ transformation is identified with the Geroch
transformation of the four-dimensional vacuum Einstein equation in terms of the
Ka\l uza--Klein mechanism.
\medskip

\medskip
\end{abstract}
]
\section{Introduction}

The subject whether Einstein's field equations have hidden symmetry or
not has been studied for a long time (see Ref.~\cite{Maison00} for a comprehensive review).
As is well known, when the four-dimensional Einstein--Maxwell field
admits a non-null Killing vector field, the system may be described in
terms of two complex scalar potentials~\cite{Ernst68b}.
The equations of motion for their potentials are derived from the
non-linear $\sigma$-model action, which is invariant under the
$SU(2,1)$ group acting~\cite{Kinnersley73}.
The invariant transformations of the $\sigma$-model action transform
a solution of the Einstein equation into other
solutions~\cite{KramerNeugebauer68}.
These transformations contain three types of
transformation:
(i) the electromagnetic duality rotation, (ii) the twisting of the Killing vector field, 
and (iii) the generation of the electromagnetic field.
The transformation (i) is just the Fitzgerald transformation in Maxwell
theory, which is still valid if coupled with gravity.
A discrete version of this was discovered by Bonnor~\cite{Bonnor54}.
The transformation (ii) mixes the norm of the Killing vector with the twist potential,
and includes the Ehlers transformation~\cite{Ehlers57} which transforms
a given static solution into its twisted version.
The transformation (iii) is the so-called Harrison transformation~\cite{Harrison68}.
This transforms given vacuum solution into its charged version.

Let us turn to the three-dimensional gravity, then consider whether
above three types of transformation do exist or not.
The duality between electric and magnetic fields is
mathematically based on the Hodge duality, and consequently fails to exist
in three dimensions.
In the static and rotationally symmetric case, however,
a kind of electricmagnetic duality is restored~\cite{Kiem97}.
The Ehlers-type transformation appears in rather wide class of solutions.
Assuming stationary and rotational symmetry,
the Einstein--Maxwell system~\cite{Clement93} or
the Einstein--Maxwell--dilaton system~\cite{Chen99} reduces to
a $\sigma$-model action which is invariant under the twisting.
This symmetry however locally results in coordinate transformations, 
so that one might not regard this as a hidden symmetry.
The subject considered here is the generation of the electromagnetic field.
We show that the Harrison-type transformation exists in the
case of the three-dimensional Einstein--Maxwell system admitting a
non-null and non-twisting Killing vector field.
In this case, a part of the field equations are derived from a
$\sigma$-model action which is invariant under some symmetry group,
and then the little group which preserves all the field equations is the
Heisenberg group.
This group includes Harrison-like transformation which generates
space-times with the Maxwell field from a locally flat space-time.
We also study the symmetry in the three-dimensional
Einstein--Maxwell--dilaton system admitting a non-null and
non-twisting Killing vector field. We show that the $SL(2,{\bf R})$
symmetry similar to the S(uperfield)-duality \cite{Sen93} exists.
For the special value of the dilaton coupling constant, this $SL(2,{\bf R})$
group transformation may be identified with the Geroch
transformation~\cite{Geroch71} of the four-dimensional stationary
vacuum space-time via the Ka\l uza--Klein method.

This paper is organized as follows.
In Sec.~\ref{II}, we analyze the symmetry of the three-dimensional
Einstein--Maxwell equations with a non-null and non-twisting Killing
vector field, where we show that the symmetry group is the Heisenberg
group.
In Sec.~\ref{III}, we turn to the Einstein--Maxwell--dilaton equations
with a non-null and non-twisting Killing vector field, and show that
the $SL(2,{\bf R})$ symmetry exists.
In Sec.~\ref{IV}, we consider the Ka\l uza--Klein dimensional reduction of
the four-dimensional vacuum space-time, and identify the
the $SL(2,{\bf R})$ symmetry of reduced system with the Geroch symmetry 
\cite{KramerNeugebauer68,Geroch71} of
the original system.
The Section \ref{V} contains the conclusion and some relevant considerations.

\section{Symmetry of the three-dimensional Einstein--Maxwell equations}
\label{II}

We consider the three-dimensional Einstein--Maxwell theory in the static case.
Let $(M^3,g)$ be a three-dimensional static space-time.
The metric can be written in the form
\begin{equation}
g=e^{2\Omega(x,y)}[-\d t^2+e^{2\Psi(x,y)}(\d x^2+\d y^2)].
\label{staticmetric}
\end{equation}
The Ricci tensor becomes
\begin{eqnarray}
&&{\rm Ric}=-e^{-2(\Psi+\Omega)}(\nabla^2\Omega+\nabla\Omega\cdot\nabla\Omega)g
-(\nabla^2\Psi+\nabla\Psi\cdot\nabla\Omega)\delta\nonumber\\
&&{}-(\nabla\otimes\nabla)\Omega+\nabla\Omega\otimes\nabla\Omega+\nabla\Omega\otimes
\nabla\Psi+\nabla\Psi\otimes\nabla\Omega,
\end{eqnarray}
where $\delta=\d x^2+\d y^2$ is the standard metric of the two-dimensional
Euclidean space $E^2$,
and the dot and $\nabla=(\partial_x,\partial_y)$
denote the inner product and the (vector) differential operator naturally defined on $E^2$,
respectively.

Next, we consider the  Maxwell equations.
The divergence equation $\d*F=0$ implies that there locally exists a
dual potential $Y$ such that
\begin{equation}
F=*\d Y,
\end{equation}
where ``$*$'' denotes the Hodge operator.
As is well known, the source-free Maxwell field
is equivalent to the minimally coupled massless scalar field in three dimensions,
if the dual potential $Y$ is identified with the scalar field.

We assume that the dual potential $Y$ is time-independent: $\partial_t Y=0$.
Then, the Bianchi identity for the Maxwell field $\d F=0$
becomes an elliptic equation on $E^2$
\begin{equation}
\nabla^2 Y+\nabla Y\cdot\nabla\Omega=0.\label{bianchi}
\end{equation}

The Einstein equation ${\rm Ric}=2\d Y\otimes\d Y$ can be written
in two-dimensional forms as
\begin{equation}
\nabla^2\Omega+\nabla\Omega\cdot\nabla\Omega=0,\label{poisson}
\end{equation}
and
\begin{eqnarray}
(&&\nabla^2\Psi+\nabla\Psi\cdot\nabla\Omega)\delta=
-(\nabla\otimes\nabla)\Omega+\nabla\Omega\otimes\nabla\Omega\nonumber\\
&&{}+\nabla\Omega\otimes\nabla\Psi
+\nabla\Psi\otimes\nabla\Omega-2\nabla Y\otimes\nabla Y.\label{tensor}
\end{eqnarray}
The trace part of the tensor equation (\ref{tensor}) becomes.
\begin{equation}
\nabla^2\Psi
=\nabla\Omega\cdot\nabla\Omega-\nabla Y\cdot\nabla Y.
\label{trace}
\end{equation}

One can easily check that a part of the Einstein--Maxwell equations
[Eqs.~(\ref{bianchi}), (\ref{poisson}) and (\ref{trace})]
are derived from the variation of the energy integral given by
\begin{equation}
I=\half\int\d x\d y e^{\Omega}[(\nabla\Omega)^2+\nabla\Omega\cdot\nabla\Psi
-(\nabla Y)^2].\label{energy}
\end{equation}
This describes a nonlinear $\sigma$-model with the base space
$(E^2,\delta)$ and the target space $(T,G)$, where the target
space metric is
\begin{equation}
G=e^{\Omega}(\d\Omega^2+\d\Omega\d\Psi-\d Y^2).
\end{equation}
In other words, a solution of the Einstein--Maxwell equations 
$\{\Omega,\Psi, Y\}$ must be a harmonic mapping:
$E^2\rightarrow T$, and
the solution space is further constrained by the trace-free part of Eq.~(\ref{tensor}):
\begin{eqnarray}
&&[(\nabla\Omega)^2+\nabla\Omega\cdot\nabla\Psi-(\nabla Y)^2]\delta
=-(\nabla\o\nabla)\Omega+\nabla\Omega\o\nabla\Omega\nonumber\\
&&
{}+\nabla\Omega\o\nabla\Psi+\nabla\Psi\o\nabla\Omega-2\nabla Y\o\nabla Y.
\label{tracefree}
\end{eqnarray}

On the other hand, a Killing vector $\xi$ of $(T,G)$:
${\mathcal L}_\xi G=0$ generates a one-parameter
group of transformation of the harmonic mappings\footnote{
More precisely, the transformation
generated by a conformal Killing vector (with a constant conformal factor) of $(T,G)$ 
leaves the $\sigma$--model action (\ref{energy}) invariant
up to a constant factor. 
}.
This means that given a solution $\{\Omega,\Psi,Y\}$ of the Einstein--Maxwell equations,
we can obtain another harmonic mapping $\{\Omega',\Psi',Y'\}$,
though the latter should satisfy Eq.~(\ref{tracefree}) to be a solution
of the Einstein--Maxwell equations.
Let us consider this transformation group that preserves the
Einstein--Maxwell equations.
It is convenient to introduce variables $U=e^{\Omega}/2$ and $V=\Omega+\Psi$.
Then the target space metric is simplified as
\begin{equation}
G=2\d U\d V-2U\d Y^2.
\end{equation}
The Killing vectors are
\begin{eqnarray}
&&\xi_1=U\partial_U-V\partial_V-\half Y\partial_{Y},\\
&&\xi_2=Y\partial_V+\half\ln (2U)\partial_Y,\\
&&\xi_3=\partial_V,\\
&&\xi_4=\partial_Y,
\end{eqnarray}
These give infinitesimal transformations of harmonic mappings.
To obtain finite transformations, we exponentiate these infinitesimal
generators $\xi$.
Taking into account Eq.~(\ref{tracefree}),
we find that the transformation group of the Einstein--Maxwell equations
are given by
\begin{eqnarray}
\label{tr1}&&\{\Omega',\Psi', Y'\}=\{\Omega,\Psi,\epsilon Y\},\\
\label{tr2}&&\{\Omega',\Psi', Y'\}=
\{\Omega,\Psi+\beta Y+\frac{\beta^2}{4}\Omega, Y+\frac{\beta}{2}\Omega\},\\
\label{tr3}&&\{\Omega',\Psi', Y'\}=\{\Omega,\Psi+\gamma, Y+\delta\},
\end{eqnarray}
where $\epsilon=\pm1$, and $\beta$, $\gamma$ and $\delta$ are real numbers.
The parameter $\epsilon$ just exchanges the sign of
the Maxwell field.
The parameter $\beta$ gives a non-trivial transformation.
In particular, $\beta$ generates non-vanishing Maxwell field
starting with a locally flat space-time.
The parameters $\gamma$ and $\delta$ result in gauge transformations.
The continuous transformations generated by parameters $\beta$, $\gamma$ and $\delta$
form the Heisenberg group (see Appendix \ref{appendixA}).

The above argument also holds when there is a hypersurface-orthogonal
 spacelike Killing vector
instead of a static Killing vector.
Then, the metric can be written as
\begin{equation}
g=e^{2\Omega(t,r)}[e^{2\Psi(t,r)}(-\d t^2+\d r^2)+\d \varphi^2],
\end{equation}
and the transformations (\ref{tr1}), (\ref{tr2}) and (\ref{tr3})
hold without modifications.

As a simple example, we show a transformation of a flat space into Einstein--Maxwell fields.
The usual Minkowski metric does not generate a non-trivial solution, however,
there is another static form of the flat space-time
\begin{equation}
g=-[\ln (r/r_0)]^2\d t^2+\left(r_1/r\right)^2(\d r^2+r^2 \d\varphi^2),
\end{equation}
where $r_0$ and $r_1$ are constants, and the null hypersurface $\{r=r_0\}$
is the Rindler horizon.
The equation (\ref{tr2}) transforms this metric into
\begin{equation}
g=-[\ln (r/r_0)]^2\d t^2+(r_1/r)^2[\ln(r/r_0)]^{\beta^2/2}(\d r^2+r^2\d\varphi^2),
\end{equation}
and generates the dual potential
\begin{equation}
Y=\frac{\beta}{2}\ln\left(\ln\frac{r}{r_0}\right).
\end{equation}
Then, the Maxwell field becomes
\begin{equation}
F=-\frac{\beta}{2}\d t\wedge\d\varphi,
\end{equation}
which describes the electric field along $\varphi$--axis.
\section{Symmetry of the Einstein--Maxwell--dilaton equations}
\label{III}

The method given in the previous section can be applied even when the
dilaton is included.
The action of the Einstein--Maxwell-dilaton system in the
Einstein frame is given by
\begin{equation}
S=\int *\left[R-4(\partial\Phi)^2-e^{-4\alpha\Phi}F^2\right],\label{emd-action}
\end{equation}
where $\alpha$ is the dilaton coupling constant.
From the Maxwell equation $\d *e^{-4\alpha\Phi}F=0$,
there exists a dual potential such that
\begin{equation}
F=e^{4\alpha\Phi}*\d Y.
\end{equation}
Then, the remaining field equations are
the Bianchi identity for the Maxwell field
\begin{equation}
\ast\d *\d Y=4\alpha g^{-1}(\d\Phi,\d Y),
\label{emd-bianchi0}
\end{equation}
the equation of motion of the dilaton,
\begin{equation}
\ast\d *\d\Phi=
-\alpha e^{4\alpha\Phi}g^{-1}(\d Y,\d Y),
\label{emd-dilaton0}
\end{equation}
and the Einstein equation 
\begin{equation}
{\rm Ric}
=4\d\Phi\otimes\d\Phi+2e^{4\alpha\Phi}\d Y\otimes\d Y.
\label{emd-einstein0}
\end{equation}

We again consider the static metric (\ref{staticmetric}), and assume the
time-independent field variables: $\partial_t Y=\partial_t \Phi=0$.
Then, the field equations become the scalar equations
\begin{eqnarray}
&&\nabla^2 Y+\nabla\Omega\cdot\nabla Y=-4\alpha\nabla\Phi\cdot\nabla Y,
\label{emd-bianchi}\\
&&\nabla^2\Phi+\nabla\Omega\cdot\nabla\Phi=\alpha e^{4\alpha\Phi}\nabla Y\cdot\nabla Y,
\label{emd-dilaton}\\
&&\nabla^2\Omega+\nabla\Omega\cdot\nabla\Omega=0,
\label{emd-poisson}
\end{eqnarray}
and the tensor equation
\begin{eqnarray}
(&&\nabla^2\Psi+\nabla\Psi\cdot\nabla\Omega)\delta\nonumber\\
&&=-(\nabla\otimes\nabla)\Omega+\nabla\Omega\otimes\nabla\Omega-4\nabla\Phi\otimes\nabla\Phi
\nonumber\\
&&{}+\nabla\Omega\otimes\nabla\Psi+\nabla\Psi\otimes\nabla\Omega
-2e^{4\alpha\Phi}\nabla Y\otimes\nabla Y,
\label{emd-tensor}
\end{eqnarray}
on $E^2$.
The trace part of the tensor equation (\ref{emd-tensor}) is
\begin{equation}
\nabla^2\Psi=\nabla\Omega\cdot\nabla\Omega
-2\nabla\Phi\cdot\nabla\Phi-e^{4\alpha\Phi}\nabla Y\cdot\nabla Y.
\label{emd-trace}
\end{equation}
The equations (\ref{emd-bianchi}), (\ref{emd-dilaton}), (\ref{emd-poisson}) and
 (\ref{emd-trace}) are derived from the variation of the energy integral
\begin{eqnarray}
&&I=\half\int \d x\d y\nonumber\\
&&\times e^{\Omega}\left[(\nabla\Omega)^2+\nabla\Omega\cdot\nabla\Psi-2(\nabla\Phi)^2
-e^{4\alpha\Phi}(\nabla Y)^2\right].
\end{eqnarray}
This describes the nonlinear $\sigma$-model with the 
target space metric
\begin{equation}
G=e^{\Omega}\left[
\d\Omega^2+\d\Omega\d\Psi-2\d\Phi^2-e^{4\alpha\Phi}\d Y^2
\right].
\end{equation}
Introducing new variables
\begin{equation}
U=e^{\Omega}/2,~~~V=\Omega+\Psi,~~~X=\frac{1}{\sqrt{2}\alpha}e^{-2\alpha\Phi},
\end{equation}
the target space metric becomes
\begin{equation}
G=2\d U\d V-\left(\frac{U}{\alpha^2}\right)\frac{\d X^2+\d Y^2}{X^2}.
\end{equation}
The Killing vectors are
\begin{eqnarray}
\xi_1 &=&2XY\partial_X-(X^2-Y^2)\partial_Y,\\
\xi_2 &=&X\partial_X+Y\partial_Y,\\
\xi_3 &=&\partial_Y,\\
\xi_4 &=&\partial_V,
\end{eqnarray}
where $\xi_A$ $(A=1,2,3)$ correspond to the Killing vectors of the Poincar\'e metric.
The finite transformations preserving the trace-free part of Eq.~(\ref{emd-tensor})
consist of
\begin{eqnarray}
&&Y'=\epsilon Y,~~~(\epsilon=\pm1)\label{emd-tr1}\\
&&\Psi'=\Psi+\gamma,~~~(\gamma\in {\bf R})\label{emd-tr2}
\end{eqnarray}
and the $SL(2,{\bf R})$ transformation
\begin{equation}
\lambda'=i\frac{a\lambda+ib}{c\lambda+id},~~~(a,b,c,d\in {\bf R},~~ad-bc=1)\label{emd-tr3}
\end{equation}
where $\lambda=-X+iY$.
This $SL(2,{\bf R})$ transformation resembles the S-duality, and turns out to be the symmetry of
the action (\ref{emd-action}) without the assumption of staticity.

\section{The $SL(2,{\bf R})$ invariance from the Ka\L uza-Klein mechanism}
\label{IV}
Here we show that the origin of the $SL(2,{\bf R})$ invariance of the Einstein--Maxwell--dilaton
system can be identified with the Geroch symmetry of the vacuum Einstein equation for the
special value of the dilaton coupling constant.

Let ($M^4,\g4$) be a four-dimensional Ricci flat space-time admitting a space-like
Killing vector $\zeta$.
The metric can be decomposed into
\begin{equation}
\g4=e^{2\alpha\Phi}g+e^{-2\alpha\Phi}(\d z+2^{1/2}A)^2,
\end{equation}
where the Killing vector is $\zeta=\partial_z$, and 
the metric $g$ and the one-form $A$ are regarded as tensor fields on the
three-dimensional manifold $M^3$.
The three-dimensional effective theory is described by the action
\begin{equation}
S=\int*\left[R-2\alpha^2(\partial\Phi)^2-e^{-4\alpha\Phi}F^2\right],
\label{kkaction}
\end{equation}
where $F=\d A$. The action coincides with Eq.~(\ref{emd-action}) when $\alpha=2^{1/2}$.

Let us consider the Einstein equation on $M^4$ for a moment.
The basic variables are
the amplitude
\begin{equation}
X=\half\g4(\zeta,\zeta)=\half e^{-2\alpha\Phi},
\end{equation}
and the twist
\begin{equation}
\omega=\star(\zeta\wedge\d\zeta)=-2^{1/2}e^{-4\alpha\Phi}*F,
\end{equation}
of the Killing vector, where ``$\star$'' denotes the Hodge operator on 
$M^4$.
From the Ricci-flatness of $M^4$, the twist is closed:
\begin{equation}
\d\omega=2\star[\zeta\wedge\hat{\rm Ric}(\zeta)]=0,
\end{equation}
so that there locally exists a twist potential $Y$:
\begin{equation}
\d Y=2^{-1/2}\omega.
\end{equation}
The complex potential is defined by
\begin{equation}
\lambda=-X+iY.
\end{equation}
In terms of this,  the Ricci flat condition of $M^4$ can be written 
in the three-dimensional form as
\begin{equation}
\ast\d\ast\d\lambda=\frac{2g^{-1}(\d\lambda,\d\lambda)}{\lambda+\lambda^*}
\label{ernst}
\end{equation}
and
\begin{equation}
{\rm Ric}=\frac{2\d\lambda\otimes\d\lambda^*}{(\lambda+\lambda^*)^2},
\label{3ricci}
\end{equation}
where ${\rm Ric}$ is the Ricci tensor on ($M^3,g$).

The Eqs.~(\ref{ernst}) and (\ref{3ricci}) are equivalent to 
Eqs.~(\ref{emd-bianchi0}), (\ref{emd-dilaton0}) and (\ref{emd-einstein0}) 
when $\alpha=2^{1/2}$, 
and these are derived from the variation of the 
effective action:
\begin{equation}
S=\int*\left[R-\frac{2g^{-1}(\d\lambda,\d\lambda^*)}{(\lambda+\lambda^*)^2}\right]
\label{action}
\end{equation}
The effective action (\ref{action}) is invariant under the so-called Geroch transformation
\cite{KramerNeugebauer68,Geroch71}
\begin{equation}
\lambda'=i\frac{a\lambda+ib}{c\lambda+id},~~~(a,b,c,d\in {\bf R},~~ad-bc=1)\label{sl(2r)},
\end{equation}
which has the same form as Eq.~(\ref{emd-tr3}).
Thus, the $SL(2,\bf{R})$ invariance of the Einstein--Maxwell--dilaton system
can be identified with the Geroch transformation of the vacuum metric
when $\alpha=2^{1/2}$.

\section{Conclusion}
\label{V}

We have discussed the hidden symmetry possessed by the three-dimensional
gravity.
It has been shown that the symmetry group of the Einstein--Maxwell
equations with a non-null and non-twisting Killing vector field is the
Heisenberg group, and this group includes the Harrison-type transformation.
In the case of the Einstein--Maxwell--dilaton system admitting a
non-null and non-twisting Killing vector field, the symmetry group is
$SL(2,{\bf R})$.
For the special value of the dilaton coupling constant ($\alpha=2^{1/2}$), this symmetry may be
identified with the Geroch symmetry of the four-dimensional vacuum
space-time admitting a spacelike Killing vector field via the Ka\l uza--Klein
mechanism.

The resulting transformations require only a single non-twisting Killing
vector field.
We can therefore apply these transformations to a wide class
of the static vacuum solutions without rotational symmetry.
For instance, we can obtain non-trivial solutions 
through the Harrison-type transformation
to static multiparticle solutions~\cite{Clement85}.
Moreover, applying the S-duality-like transformation, we can have a
class of static multiparticle solutions with the non-trivial Maxwell and  dilaton fields.

Our results rely on the $\sigma$-model structure of the system,
however, the non-vanishing consmological term brakes
the $\sigma$-model structure; namely, 
the method presented here is no longer valid in the presence of the cosmological constant.
In particular, the solution generating method could not be applied to
three-dimensional black holes, which require the negative
cosmological term~\cite{Ida00}.

\section*{Acknowledgments}
We would like to thank Profs. H.~Sato, A.~Hosoya and K.~Nakao for many useful suggestions
and comments.n
We also thank Dr. A.~Ishibashi for stimulating discussions.
Y.M. would like to thank Prof. Takashi~Nakamura for continuing
encouragement.
D.I. was supported by JSPS Research and
this research was supported in part by the Grant-in-Aid for Scientific
Research Fund (No. 4318).

\appendix

\section{Group structure of finite transformation of Einstein--Maxwell 
equations}
\label{appendixA}
Let $\Gamma$ be the connected component including the unit element of the
symmetry group of Einstein--Maxwell equations. [Namely, $\Gamma$ is the set
of finite transformations except the Maxwell field inversion (\ref{tr1}).]
Each element of $\Gamma$ is uniquely represented as
\begin{equation}
f_{\beta\delta\gamma}:=\exp(\beta\xi_2+\delta\xi_4-\gamma\xi_3),
\quad \beta,\delta,\gamma\in{\bf R}.
\end{equation}
Using the Campbell--Hausdorff formula, one can obtain the product
formula
\begin{equation}
f_{\beta\delta\gamma}\circ f_{\beta'\delta'\gamma'}
=
f_{\beta+\beta',\delta+\delta',\gamma+\gamma'
+(\beta\delta'-\beta'\delta)/2},
\label{product}
\end{equation}
and the identity:
\begin{equation}
f_{\beta\delta\gamma}
=
e^{-\gamma\xi_3}e^{\delta\xi_4/2}e^{\beta\xi_2}e^{\delta\xi_4/2}
=
e^{\delta\xi_4/2}e^{\beta\xi_2}e^{\delta\xi_4/2}e^{-\gamma\xi_3}.
\label{identity}
\end{equation}
Consider a map $\phi:\Gamma\to GL(3,{\bf R})$ such that
\begin{equation}
\phi(f_{\beta\delta\gamma})=\left(
\begin{array}{ccc}
1 & \beta & \gamma+\beta\delta/2\\
0 & 1 & \delta\\
0 & 0 & 1\\
\end{array}
\right).
\end{equation}
From Eq.~(\ref{product}), it can easily be checked that $\phi$ is one-to-one
group homeomorphism.
Since the image of $\phi$ is upper triangular matrices whose diagonal elements
are unity, $\Gamma$ is the Heisenberg group so called.
From the identity (\ref{identity}) and the transformation laws
(\ref{tr2}) and (\ref{tr3}), one can calculate the group action
\begin{eqnarray}
&&
f_{\beta\delta\gamma}(\{\Omega,\Psi,Y\})
\nonumber\\&&\quad =
\left\{
\Omega,\Psi+\beta Y+\frac{\beta^2}{4}\Omega-\gamma+\frac{\beta\delta}{2},Y+\frac{\beta}{2}\Omega+\delta
\right\}.
\label{groupaction}
\end{eqnarray}
Using Eq.~(\ref{groupaction}) again, one obtains
\begin{eqnarray}
&&
f_{\beta\delta\gamma}(f_{\beta'\delta'\gamma'}(\{\Omega,\Psi,Y\}))
\nonumber\\&&\quad =
(f_{\beta'\delta'\gamma'}\circ f_{\beta,\delta,\gamma})(\{\Omega,\Psi,Y\}).
\end{eqnarray}
Therefore, $\Gamma$ is the right transformation group on the solution
space $\{\Omega,\Psi,Y\}$.

\end{document}